\documentclass{iopconfser}
\usepackage{amsmath}
\usepackage{amssymb}
\usepackage{mathrsfs}
\usepackage[pdftex,hidelinks]{hyperref}
\usepackage{cleveref}
\usepackage{orcidlink}

\DeclareSymbolFont{matha}{OML}{txmi}{m}{it}% txfonts
\DeclareMathSymbol{v}{\mathord}{matha}{118}

\newcommand{\heq}{\overset{\mathcal H}{=}}

\begin{document}

\title{Generalised focusing theorem and dynamical horizon entropy in diffeomorphism-invariant theories}

\author{Zihan~Yan\,\orcidlink{0000-0003-2998-7040}$^{1}$}

\affil{$^1$DAMTP, Centre for Mathematical Sciences, University of Cambridge,\\~\,Wilberforce Road, Cambridge, U.K. CB3 0WA}

\email{zy286@cam.ac.uk}

\begin{abstract}
I summarise recent progress on light-ray focusing and horizon thermodynamics in general diffeomorphism-invariant theories of gravity coupled to bosonic matter. In pure gravity and with scalar or vector fields, the null-null gravitational equation of motion on a linearly perturbed Killing horizon generalises the Raychaudhuri equation, defining a generalised expansion that never increases under the null energy condition. This proves a generalised focusing theorem and defines an increasing horizon entropy (Wall entropy). When higher-spin fields are present, the generalised focusing theorem persists subject to a ``higher-spin focusing condition'', which I propose as a physical consistency constraint on higher-spin theories.
\end{abstract}

\section{Introduction}\label{sec:intro}

The focusing theorem and the Bekenstein-Hawking (BH) horizon entropy are two flagship concepts in General Relativity (GR): one states that light rays bend towards each other under positive energy, revealing the attractive nature of gravity, while the other counts the microscopic degrees of freedom hidden behind a horizon, providing clues to the fundamental constituents of space and time.

When higher-curvature corrections due to quantum or stringy effects are present, the geometrical focusing theorem no longer holds, and the Bekenstein-Hawking entropy formula needs to be corrected. Some important and interesting questions immediately arise for these theories:  Is gravity still ``attractive''? Are there trapped regions and singularities? What is the correct entropy formula for (dynamical) horizons? 

In this paper based on my talk given at the GR24-Amaldi16 conference in July 2025 in Glasgow, I review some recent results of A.~Wall and myself \cite{Wall:2024lbd,Yan:2024gbz} showing how the focusing theorem is generalised to arbitrary theories of gravity coupled to bosonic matter, how a correct entropy formula (Wall entropy) for dynamically perturbed Killing horizon is defined in these theories, and how the requirement of light-ray focusing acts as a physical consistency condition for higher-spin theories. The entire research program adopts a systematic approach by considering a general class of diffeomorphism-invariant Lagrangians that encompasses most of the interesting models of effective field theories of quantum gravity. 

\section{Light-ray focusing and horizon entropy in GR}\label{sec:GR}

For a null geodesic congruence (a family of nearby geodesics), one can describe its evolution using the \emph{Raychaudhuri equation} \cite{PhysRev.98.1123}:
\begin{equation}
    \partial_v \theta = - \frac{\theta^2}{D-2} - \sigma_{ab} \sigma^{ab} - R_{ab}k^a k^b
\end{equation}
where $D$ is the spacetime dimension, $v$ is an affine null parameter with $k=\partial_v$ the null tangent vector, $\theta$ is the geometric expansion of the congruence, $\sigma_{ab}$ is the shear, and $R_{ab}$ is the Ricci tensor. Particularly, the expansion can be expressed as $\theta = \mathscr A^{-1} \partial_v \mathscr A$, where $\mathscr A$ is the infinitesimal codimension-2 area element spanned by a fixed number of nearby null geodesics. This area element provides a geometrical measure of how closely the geodesics are spaced.

The \emph{focusing theorem} in GR can be proved by using the null-null component of the Einstein equation $R_{ab} k^a k^b = 8 \pi G T_{ab} k^a k^b$---where $G$ is Newton's constant,  $T_{ab}$ is the stress-energy tensor---and the null energy condition (NEC) $T_{ab} k^a k^b \geq 0$. Plugging these into the Raychaudhuri equation, we obtain $\partial_v \theta \leq 0$, i.e., the expansion is always non-increasing, hence, light rays always tend to focus under NEC. The focusing theorem is a manifestation of the attractiveness of gravity, and it
underlies several important results in GR, such as the Penrose singularity theorem and the Hawking
area theorem.

On the other hand, in GR, the entropy of a horizon $\mathcal H$ is given by the Bekenstein-Hawking formula $S_\text{BH} = \text{Area}[\mathcal H]/4G$, i.e., the entropy is proportional to the horizon area. Such identification has two profound implications: (1) The entropy density can be viewed as a measure of proximity between nearby null geodesics; (2) The focusing theorem actually encodes the thermodynamics of near-stationary horizons. Integrating the Raychaudhuri equation over a horizon cross-section, we obtain $\partial_v^2 S_\text{BH} \leq 0$ under NEC. Together with the future horizon boundary condition $\partial_v S_\text{BH} \to 0$ as $v\to\infty$, we obtain the second law $\partial_v S_\text{BH} \geq 0$. The physical process first law can also be obtained by integrating the Raychaudhuri equation over the entire horizon.

\section{Light-ray focusing in diffeomorphism-invariant theories}\label{sec:focusing-DIT}
For general theories of gravity, presumably GR with higher-curvature corrections, the focusing theorem fails to hold. The gravitational equation of motion is $E_{ab} = (8 \pi G)^{-1} G_{ab} + H_{ab} = T_{ab}$ where $G_{ab} = R_{ab} - \frac{1}{2}g_{ab} R$ (with $g_{ab}$ the metric and $R$ the Ricci scalar) is the Einstein tensor, and $H_{ab}$ represents higher-curvature correction terms. In general, $H_{ab} k^a k^b$ has an indefinite sign. Therefore, when the equation of motion is plugged into the Raychaudhuri equation
\begin{equation}
    \partial_v \theta = - \frac{\theta^2}{D-2} - \sigma_{ab} \sigma^{ab} - 8\pi G T_{ab}k^a k^b + 8\pi G H_{ab}k^a k^b \lessgtr 0,
\end{equation}
we find $\partial_v \theta$ can take either sign. This means light rays no longer focus geometrically, even when NEC is satisfied. This is expected, however, because dynamics is no longer compatible with the geometry (kinematics). After all, $E_{ab}$ has no special geometrical meaning. 

We need a new concept of ``focusing'' if we believe that positive energy density should cause light to bend towards each other in any theory---we expect gravity to remain attractive in arbitrary theories, although it is no longer described by geometry. We want to define a \emph{generalised expansion} $\Theta$ such that it monotonically decreases under NEC, i.e., we want a ``generalised Raychaudhuri equation''
\begin{equation}
    \partial_v \Theta \propto - T_{ab} k^a k^b - (\text{quadratic terms}) \leq 0.
\end{equation}
Furthermore, this generalised expansion $\Theta$ should itself be a rate of change of some density $\varsigma$, which is a new measure of proximity between nearby light rays. It is too difficult to find such a $\Theta$ for arbitrary theories without further assumptions. Remarkably, when a background Killing horizon is present, the symmetry allows us to uniquely define $\Theta$ on the horizon to linear order in perturbations for any theory.

In diffeomorphism-invariant theories (DIT) of gravity and non-minimally coupled scalar and vector fields $\phi^I, V_a^J$ with Lagrangian
\begin{equation}
    L(g^{ab},R_{abcd},\phi^I,V^J_a,\nabla_a),
\end{equation}
(where $R_{abcd}$ is the Riemann tensor), the null-null component of the \emph{off-shell} gravitational equation of motion admits a generalised expansion on a linearly perturbed Killing horizon $\mathcal H$:
\begin{equation}
    2 \pi E_{ab} k^a k^b \heq - \partial_v \Theta + \mathcal O(\epsilon^2)
\end{equation}
valid at the linear order of the perturbation parameter $\epsilon$. We call this the \emph{generalised linear Raychaudhuri equation} (GLRE). The generalised expansion $\Theta$ has the structure
\begin{equation}
    \Theta = \partial_v \varsigma + D_i J^i
\end{equation}
where $\varsigma$ is the ``entropy'' density, and $J^i$ is the ``entropy'' current (unimportant for compact horizons) \cite{Wall:2024lbd,Wall:2015raa,Bhattacharya:2019qal,Bhattacharyya:2021jhr,Biswas:2022grc,Deo:2023vvb}. The reason for the naming will be clear in \cref{sec:entropy-DIT}. 

Consider perturbations that are sourced by some external minimally coupled test matter field with stress-energy tensor $T_{ab} \sim \mathcal O(\epsilon)$ that obeys the NEC (i.e., $T_{ab} k^a k^b \geq 0$). Plugging the perturbed equation of motion $E_{ab} = T_{ab}$ into the GLRE, we prove the \emph{generalised focusing theorem} on the horizon:
\begin{equation}
    \partial_v \Theta \heq - 2\pi T_{ab} k^a k^b \leq 0, \label{eq:focusing-eq}
\end{equation}
i.e., the generalised expansion $\Theta$ never increases. Light rays tend to converge if their proximity is measured in the $(\varsigma,J^i)$ ``entropy'' density-current. This echoes implications (1) and (2) in \cref{sec:GR} and suggests that the Riemannian geometry in GR should be replaced by some ``entropic geometry'' in DIT when studying the evolution of null geodesic congruences.

\section{Entropy of non-stationary horizons in diffeomorphism-invariant theories}\label{sec:entropy-DIT}
We had named $(\varsigma,J^i)$ the ``entropy'' density-current. How do we know it underlies an entropy? Consider a compact cross-section $\mathcal C$ of a horizon $\mathcal H$. The codimension-2 integral of $\varsigma$ over $\mathcal{C}$ 
\begin{equation}
    S_\text{Wall} = \int_\mathcal{C} \varsigma\, \mathrm dA
\end{equation}
is the proposed horizon entropy. It is known as the \emph{Wall entropy}, originally defined by A.~Wall in \cite{Wall:2015raa} for pure gravity and gravity with scalar fields. It is well-defined as it satisfies thermodynamic laws:

\noindent\textbf{First Law.} Multiply \cref{eq:focusing-eq} by $\kappa v$ and integrate over $\mathcal H$, we have the physical process first law for $S_\text{Wall}$
\begin{equation}
    \frac{\kappa}{2 \pi} \Delta S_{\text{Wall}} = \int_{0}^{\infty} \mathrm d v\int_{\mathcal C(v)}\mathrm dA~ T_{ab}k^a \xi^b = \Delta M - \Omega_{\mathcal H}\Delta J
\end{equation}
where $\xi^a \heq \kappa v k^a$ is the background Killing vector, $\kappa$ is the surface gravity, $\Delta$ labels the change between $v=\infty$ and $v=0$, and $M,J$ are the energy and angular momentum associated to the horizon.

\noindent\textbf{Second Law.} Integrating \cref{eq:focusing-eq} over $\mathcal{C}(v)$ at any affine parameter $v$, we have $\partial_v^2 S_\text{Wall} \leq 0$. Together with the future horizon boundary condition $\partial_v S_\text{Wall} \to 0$ as $v \to \infty$ (the horizon settles down long after perturbations), we obtain the second law 
\begin{equation}
    \partial_v S_\text{Wall} \geq 0.
\end{equation}
There are more reasons to trust Wall entropy: (a) The Wall entropy density-current correctly restores the bending of light under NEC. (b) It is a codimension-2 integral of local geometrical quantities. (c) It is unique and does not suffer from Jacobson-Kang-Myers ambiguities \cite{Wall:2024lbd,Wall:2015raa,Iyer:1994ys,Jacobson:1993vj}. (d) It reduces to BH entropy in GR. It reduces to Wald entropy \cite{Wald:1993nt} for stationary horizons. (e) It is covariant and independent of the choice of Gaussian null coordinates \cite{Hollands:2022fkn}. (f) For $f(\text{Riemann})$ theory, it is identical to the holographic entanglement entropy (Dong entropy) \cite{Dong:2013qoa} although the full equivalence is still a conjecture.

Wall entropy is closely related to another dynamical generalisation of Wald entropy---the Holland-Wald-Zhang (HWZ) dynamical entropy \cite{Hollands:2024vbe,Visser:2024pwz}, which is defined by the improved Noether charge of the background Killing symmetry. The two entropy formulae are related by $S_\text{HWZ} = (1-v\partial_v) S_\text{Wall}$. In GR, we further identify $S_\text{HWZ}[\mathcal H] = S_\text{Wall}[\mathcal A]$, where $\mathcal A$ is the apparent horizon that lies perturbatively away from the dynamical event horizon $\mathcal H$. In DIT, one can define a \emph{generalised apparent horizon} $\mathcal A_\text{gen}$ by extending $\Theta$ off the event horizon and demanding $\Theta = 0$, and one can show $S_\text{HWZ}[\mathcal H] = S_\text{Wall}[\mathcal A_\text{gen}]$ \cite{HVWYZ}.

\section{Constraints from light-ray focusing on higher-spin fields}\label{sec:higher-spin}
When higher-spin fields $\phi_{a_1 \cdots a_s}$ with $s\geq 2$ (except massless graviton) are present in the DIT, the off-shell null-null equation of motion on the perturbed Killing horizon becomes \cite{Yan:2024gbz}
\begin{equation}
    2 \pi E_{ab} k^a k^b \heq - \partial_v \Theta - \mathcal L_\xi \mathcal P_2.
\end{equation}
There is an \emph{indefinite term} $\mathcal L_\xi\mathcal P_2$ (vanishes for stationary horizons) which cannot be absorbed into $\Theta$ but can take either sign. This indefinite term breaks the structure of GLRE. Even if the external perturbing source $T_{ab}$ satisfies NEC, in general, $\Theta$ is not a well-defined generalised expansion, and it makes no sense to discuss focusing, entropy and second law, etc. This is also expected since an arbitrary theory involving higher-spin fields is most likely to be unphysical due to violations of causality, unitarity, etc. In this case, we use the focusing of light as a physical constraint on higher-spin theories. We demand $\mathcal L_\xi \mathcal P_2 = 0$, which guarantees the generalised focusing theorem to hold and the Wall entropy to be well-defined as before. We call this the \emph{higher-spin focusing condition} \cite{Yan:2024gbz}.

The indefinite term $\mathcal L_\xi \mathcal P_2$ is a linear function of the perturbed higher-spin field components $\delta \phi_{(w\geq2)}$ of boost weight $w\geq 2$, e.g., for symmetric $\phi_{a_1 \cdots a_s}$, these consist of $\delta \phi_{vv}$ for spin-2, $\delta \phi_{vvv}, \delta \phi_{vvi}$ for spin-3, $\delta \phi_{vvvv}, \delta \phi_{vvvi}, \delta \phi_{vvij},\delta \phi_{vvvu}$ for spin-4, and so on. Here, $i,j$ labels the codimension-2 spatial directions and $u$ labels the ingoing null direction. One way to satisfy the focusing condition is that the higher-spin theory possesses enough gauge symmetry to gauge away $\delta \phi_{(w\geq2)}$ near the horizon. This is known as the \emph{gauge type} focusing condition \cite{Yan:2024gbz}.

As an example, the non-rotating black hole solution in 3d $\mathfrak{sl}(3,\mathbb R) \oplus \mathfrak{sl}(3,\mathbb R)$ higher-spin gravity with a spin-3 field $\varphi_{(abc)}$ \cite{Gutperle:2011kf} satisfies the focusing condition. In the black hole gauge, near the horizon the solution reads
\begin{align}
    \mathrm ds^2 &\approx - \kappa^2 r^2 \,\mathrm dt^2 + \mathrm dr^2 + g_{\phi\phi}(0)\,\mathrm d\phi^2\\
    \varphi_{abc} \,\mathrm dx^a \mathrm dx^b \mathrm dx^c & = \varphi_{\phi rr}\,\mathrm d \phi \,\mathrm d r^2 + \varphi_{\phi tt}\,\mathrm d\phi\, \mathrm dt^2 + \varphi_{\phi \phi \phi} \,\mathrm d \phi^3.
\end{align}
Transforming to Gaussian null coordinates $(u,v,\phi)$ with $v$ the affine null parameter of horizon generators, we have $\varphi_{vvv}=0$, $\varphi_{vv\phi}|_\mathcal{H} = 0$, making $\mathcal L_\xi \mathcal P_2 = 0$. In this example, the higher-spin focusing condition is satisfied, thus the generalised focusing theorem holds, and the Wall entropy is well-defined.

\section{Outlook}\label{sec:outlook}
For future work, we aim to generalise the results above to general theories of gravity coupled to quantum fields to prove a quantum focusing theorem, which would imply a generalised second law, a quantum singularity theorem, and a quantum Bousso bound in general DIT. We also intend to work beyond the linear order of perturbation to obtain a generalised non-linear Raychaudhuri equation and a non-perturbative focusing theorem for effective field theories of quantum gravity.

\section*{Acknowledgements}
I am grateful to Raphael~Bousso, Alejandra~Castro, Prateksh~Dhivakar, Zucheng~Gao, Gary~Horowitz, Veronika~Hubeny, Zhihan~Liu, Prahar~Mitra, Harvey~Reall, Jorge~Santos, Arvin~Shahbazi-Moghaddam, Ronak~Soni, Manus~Visser, Robert~Wald, Aron~Wall, Diandian~Wang, Zi-Yue~Wang, Houwen~Wu, and Victor~Zhang for helpful discussions. I especially thank Aron~Wall for extensive and insightful conversations. This work is supported by an Internal Graduate Studentship of Trinity College, Cambridge, AFOSR grant FA9550-19-1-0260 ``Tensor Networks and Holographic Spacetime'' and grant NSF PHY-2309135 to the Kavli Institute for Theoretical Physics (KITP).

\bibliography{references.bib}

\end{document}